\normalfont%
\documentclass[%
 aip,
 amsmath,amssymb,
 reprint,floatfix,
groupedaddress,%
]{revtex4-1}

\usepackage{graphicx}
\usepackage{dcolumn}
\usepackage{bm}
\usepackage{verbatim} 
\usepackage{gensymb} 
\usepackage{epstopdf} 
\usepackage[english]{babel}

\begin{document}

\preprint{AIP/123-QED}

\title{Ball-grid array architecture for microfabricated ion traps}

\author{Nicholas D. Guise}
\email{nicholas.guise@gtri.gatech.edu}
\author{Spencer D. Fallek}
\author{Kelly E. Stevens}
\author{K. R. Brown}
\author{Curtis Volin}
\author{Alexa W. Harter}
\author{Jason M. Amini}
\affiliation{Georgia Tech Research Institute,  Atlanta, Georgia 30332, USA}

\author{Robert E. Higashi, Son Thai Lu, Helen M. Chanhvongsak, Thi A. Nguyen, Matthew S. Marcus, Thomas R. Ohnstein, and Daniel W. Youngner}
\affiliation{Honeywell International, Golden Valley, Minnesota 55422, USA}

\date{\today}

\begin{abstract}
State-of-the-art microfabricated ion traps for quantum information research are approaching nearly one hundred control electrodes. We report here on the development and testing of a new architecture for microfabricated ion traps, built around ball-grid array (BGA) connections, that is suitable for increasingly complex trap designs. In the BGA trap, through-substrate vias bring electrical signals from the back side of the trap die to the surface trap structure on the top side. Gold-ball bump bonds connect the back side of the trap die to an interposer for signal routing from the carrier.  Trench capacitors fabricated into the trap die replace area-intensive surface or edge capacitors. Wirebonds in the BGA architecture are moved to the interposer. These last two features allow the trap die to be reduced to only the area required to produce trapping fields. The smaller trap dimensions allow tight focusing of an addressing laser beam for fast single-qubit rotations. Performance of the BGA trap as characterized with $^{40}$Ca$^+$ ions is comparable to previous surface-electrode traps in terms of ion heating rate, mode frequency stability, and storage lifetime. We demonstrate two-qubit entanglement operations with $^{171}$Yb$^+$ ions in a second BGA trap.

\end{abstract}

\maketitle

\section{Introduction}

Trapped atomic ions are a leading platform in quantum information research, offering the advantages of precise optical manipulation and long qubit coherence times. The transport architecture proposed for scalable trapped-ion quantum computing \cite{Kielpinski2002} requires a complex system in which large numbers of ions are shuttled to various locations in the trap structure. Towards this goal, microfabrication techniques have enabled miniaturization of ion traps \cite{Chiaverini2005} that allow for the incorporation of arrays of microfabricated electrodes on a planar substrate (see, for example, Refs. \cite{Stick2006,Seidelin2006,Schulz2008,Allcock2010,Brady2011,Moehring2011,Wright2013}). In the typical trap, surface electrode structures are fabricated on a thin chip that sits atop a commercial ceramic pin-grid array (CPGA) carrier.  To supply DC trapping potentials for each electrode, pads on the CPGA are wirebonded to pads on the trap chip. Capacitors on the surface of the chip or bonded at the perimeter of the carrier are used to filter out radio frequency (RF) pickup on the electrodes.

\begin{figure}[htbp]
\includegraphics[width=\columnwidth]{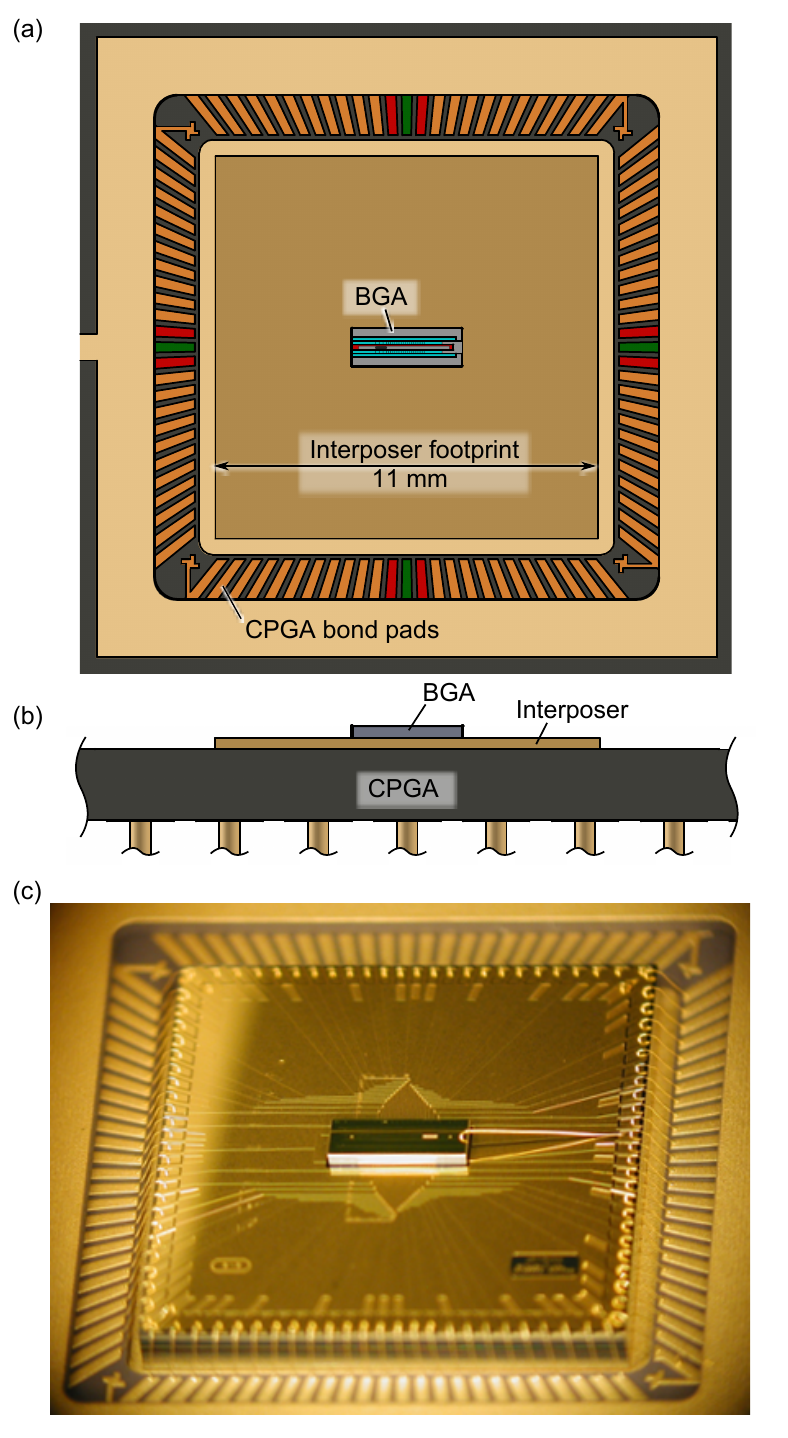}
\caption{\label{fig:Overview} Overview of the BGA design: (a) Die bond region of the CPGA showing the BGA trap and the interposer footprint. (b) Side view.  (c) Fully packaged BGA trap. The long bond wire supplies the trap RF signal.}
\end{figure}

Existing fabrication techniques will limit scaling surface-electrode ion traps far beyond those in use today.  Recent trap designs have grown to incorporate nearly 100 DC control electrodes \cite{Wright2013}.  For linear traps with as few as 50 electrodes \cite{Doret2012}, the capacitors and bond pads can consume a majority of the overall trap chip area and perimeter, strongly constraining the layout of electrode structures and DC/RF lead traces.  Wirebonds must be carefully arranged to minimize obstructions to laser access, as beams for trapping and qubit manipulation must interact with ions confined $\approx 100$ $\mu$m above the chip surface.  The large periphery (non-trap area) of the trap die presents geometric limitations to focusing these lasers onto the ion while maintaining the narrow beam waists preferred for operations such as addressing individual ions in a chain or driving hyperfine Raman transitions \cite{Rohde2001,Ozeri2007,Choi2014}. For example, a 729 nm beam for single qubit addressing, focused to a waist of 3.4 $\mu$m as in Ref. \cite{Rohde2001}, has a Rayleigh range of 0.05 mm and would diverge to a beam radius of 100 $\mu$m within 1.5 mm, incompatible with existing traditional surface trap dimensions.  Novel trap architectures remove portions of the trap in several dimensions but are still constrained by wirebonds along two edges \cite{Maunz2013DAMOP}.

\begin{figure*}[htbp]
\includegraphics[width=\textwidth]{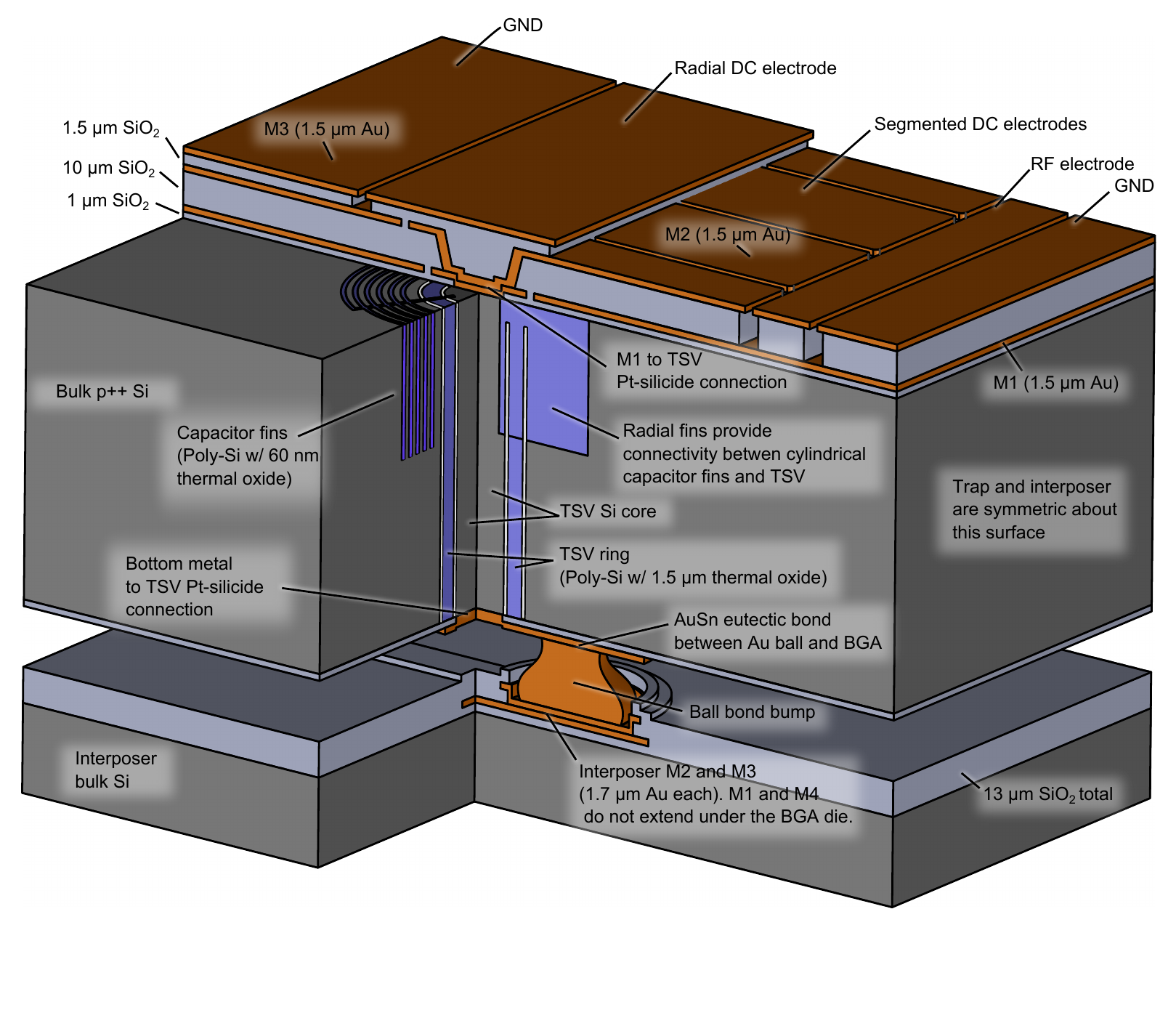}
\caption{\label{fig:CrossSection} Schematic cross section through a BGA trap die and interposer (not to scale).}
\end{figure*}

We demonstrate here a new architecture for microfabricated ion traps, built around ball-grid array (BGA) connections.  Our BGA design (Fig. \ref{fig:Overview}) mitigates scalability concerns related to the capacitors and wirebond connections, providing a flexible architecture for traps of increased complexity.  Electrodes in the BGA trap are connected with through-substrate vias (TSVs) to pads on the back side of the trap die.  Gold-ball bump bonds then connect these back-side pads to traces on an interposer.  The interposer is wirebonded to a CPGA carrier for signal routing. The 100-pin Kyocera CPGA used here is compatible with our existing test setups; alternate carriers would require only a redesign of the relatively simple interposer, rather than modifications to the trap chip itself. Trench capacitors are fabricated into the trap die, similar to a proposal in Ref. \cite{Allcock2012}, reducing the trap die area by a factor of thirty over traps with planar capacitors. Filter capacitors located at the edge of the CPGA would be less effective in this architecture, due to distance from the trap. With the wirebond connections relocated to the interposer, the surface of the BGA trap chip remains unobstructed for laser access. Gold is utilized for the electrode surfaces, minimizing oxide formation which is a potential source of charging and stray fields in surface electrode traps. A re-entrant metallization profile screens any trapped charge that might accrue on dielectric surfaces \cite{Harlander2010,Allcock2012}. Details of the BGA trap fabrication process are provided in Section \ref{sec:Fabrication}.

We characterize the BGA trap by loading single and multiple $^{40}$Ca$^+$ ions.  Performance of the trap is comparable to previous microfabricated surface traps in terms of ion heating rate, axial mode stability, and storage lifetime for one and two trapped ions \cite{Daniilidis2011,Doret2012,Guise2014,Graham2014}.  We take advantage of the reduced trap die size and improved optical access to demonstrate tighter focusing of a 729 nm laser beam with resulting speedup in single-qubit rotation rates. Results from the BGA ion trap testing are presented in Section \ref{sec:Testing}.  We present in Section \ref{sec:MSGate} a demonstration of two-qubit entanglement with $^{171}$Yb$^+$ ions in a second BGA trap. 

\section{\label{sec:Fabrication}Fabrication}
\subsection{Overview}

The BGA trap includes 48 trench capacitors and 48 TSVs connected to a total of 48 DC electrodes.  A schematic cross section of the BGA trap is shown in Figure \ref{fig:CrossSection}, highlighting the metal and insulator layers, trench capacitors, and TSVs.  The trap die is bonded to an interposer (Sec. \ref{sec:Interposer}), which serves to elevate the trap above the carrier for improved laser access and to route electrical connections from the package to the trap die.

\subsection{Processing Details}
\subsubsection{\label{sec:TrapFab}Trap Die Fabrication}

The process for making the trap die begins with a 500 $\mu$m-thick p++ (heavily boron-doped) silicon wafer with a resistivity of $0.001-0.005$ $\Omega\cdot$cm. The key fabrication steps are described below.

\begin{figure}[htbp]
\includegraphics[width=\columnwidth]{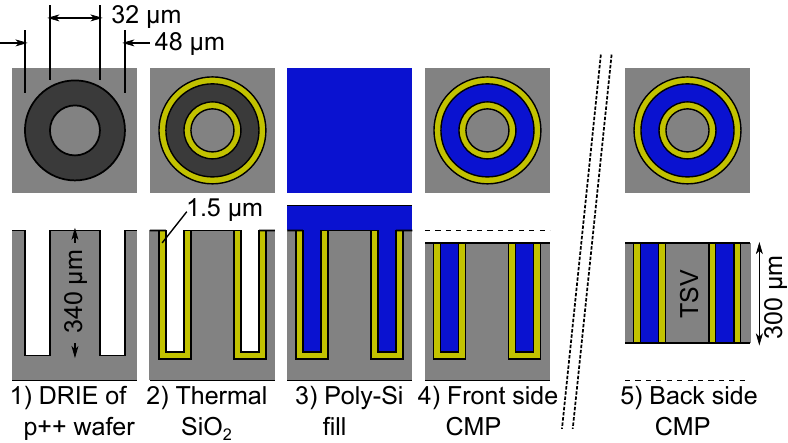}
\caption{\label{fig:TSVProcess} Process for making TSVs.}
\end{figure}

\begin{figure}[htbp]
\includegraphics[width=\columnwidth]{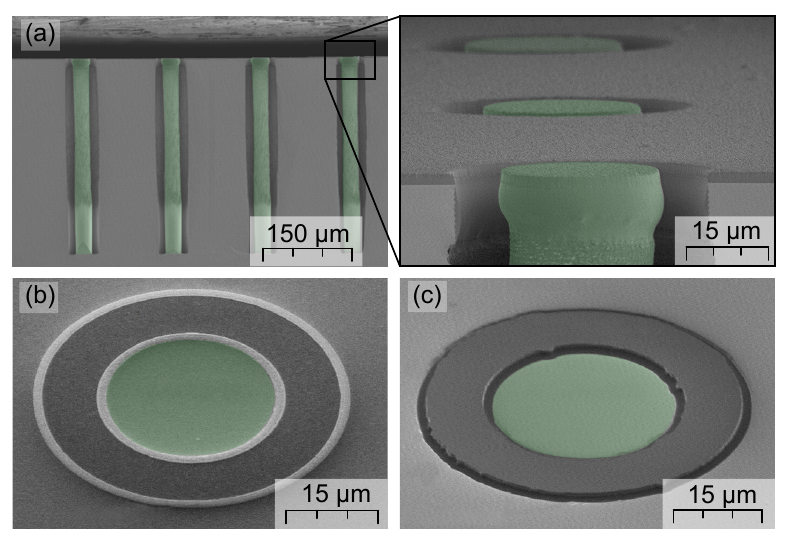}
\caption{\label{fig:TSVSEMs} SEM images of the TSVs. The bulk Si core has been highlighted in green. (a) cross-section and top detail after DRIE and before the first thermal oxidation; (b) bottom view after back-side CMP; (c) high-angle surface view before thermal oxidation.}
\end{figure}

(i) Making TSVs: Figure \ref{fig:TSVProcess} shows the complete process for forming the TSVs. Scanning electron microscope (SEM) images of the TSV features are shown in Fig. \ref{fig:TSVSEMs}. The outlines for the TSVs are patterned with ring-shaped features and etched $\approx 340$ $\mu$m into the wafer using a Deep Reactive Ion Etch (DRIE) tool (Fig. \ref{fig:TSVProcess}, step 1).  The side and bottom walls of the TSVs are oxidized in a thermal oxidation furnace to a thickness of 1.5 $\mu$m (step 2).  The trenches are filled with $\approx 6$ $\mu$m of highly conductive p++ polysilicon, completely sealing the tops of the holes (step 3). A Chemical Mechanical Polish (CMP) is performed from the front side to expose the original bulk silicon wafer (step 4).  Much later in the fabrication process (after front-side metallization), a back-side CMP operation (step 5) reduces the wafer thickness to 300 $\mu$m and exposes the TSV ring. The TSV resistances are $40 \pm 2 \; \Omega$, with breakdown voltages in excess of 380 V, above the range of the test equipment used.

\begin{figure}[htbp]
\includegraphics[width=\columnwidth]{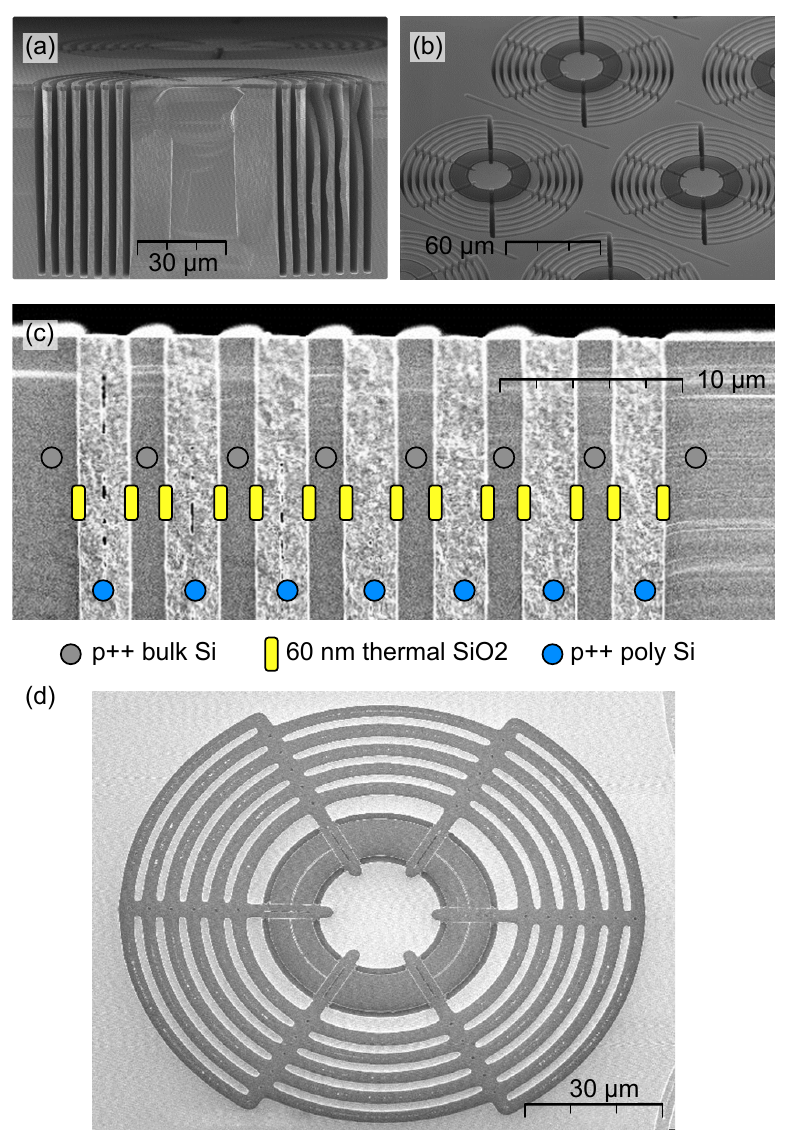}
\caption{\label{fig:TrenchCapSEMs} SEM images of the trench capacitors: (a) cross-section view and (b) surface view after DRIE trench capacitor etch; (c) cross-section view and (d) surface view after polysilicon fill and CMP.}
\end{figure}

(ii) Making Trench Capacitors: Fabrication of the trench capacitors begins with DRIE of trenches arranged in concentric rings around the TSVs, 55-70 $\mu$m into the silicon wafer.  Thermal oxide of thickness 60 nm is grown on the sidewalls of the trenches.  The trenches are then filled with p++ polysilicon.  CMP is performed on the wafer surface until the bulk silicon wafer is exposed.  Finally, 0.5 $\mu$m of thermal SiO$_2$ is grown to consume CMP damage. Figure \ref{fig:TrenchCapSEMs} shows SEM images at various stages of trench capacitor fabrication.  At the end of trench capacitor processing (Fig. \ref{fig:TrenchCapSEMs}d), the trench capacitors remain electrically isolated from the TSVs by a thin oxide layer.

We measure capacitance values of $103.5 \pm 1.2$ pF for the 48 trench capacitors on a trap die.  Capacitance values in the final packaged trap increase to $118 \pm 3$ pF, with variation dominated by differences in the lengths and widths of lead traces and electrodes.  Typical breakdown voltage for a trench capacitor is $\approx 18$ V, determined by testing 100 capacitors on a test wafer.

(iii) Making Ohmic Contacts to TSVs and Trench Capacitors: Once the trench capacitors and TSVs have been formed, a 1.0 $\mu$m layer of chemical vapor deposition (CVD) SiO$_2$ is deposited on top of the wafers. A layer of photoresist is patterned connecting the TSV cores and radial arms of the trench capacitors.  The wafers are then etched using a RIE (Reactive Ion Etch) process.  Once the underlying silicon has been reached, we form a platinum silicide layer on the tops of the trench capacitors and TSV cores.  The platinum silicide provides Ohmic contacts for electrical connections between the doped silicon structures (TSVs and associated trench capacitors) and the front-side metallization layers described below. 

(iv) Patterning Metal Layers: The three front-side metallization and dielectric layers are shown schematically in Fig. \ref{fig:CrossSection}. The bottom most metal layer (M1) is essentially a large ground plane, interrupted only where there are TSVs. The second metal layer (M2) forms the trap DC and RF electrodes. A 10 $\mu$m oxide layer between M1 and M2 limits the capacitance of the RF electrodes. We utilize a third metal layer (M3) so that the M1 to TSV connection can be planarized with oxide, which prevents topography caused by that connection from affecting the ion trapping potentials. M3 contains the radial DC electrodes and a ground plane. Appendix \ref{app:Metallization} provides additional fabrication details of the front-side metallization.

After M3 is patterned, a $>2$ hour RIE step anisotropically removes exposed SiO$_2$ by etching in the vertical direction. A 3 $\mu$m BOE (Buffered Oxide Etch) process then etches the SiO$_2$ laterally, leaving overhanging metal features which aid in screening dielectric from laser light during trap operation and which shield the ion to some extent from stray fields caused by trapped charge on the dielectrics.

\begin{figure}[htbp]
\includegraphics[width=\columnwidth]{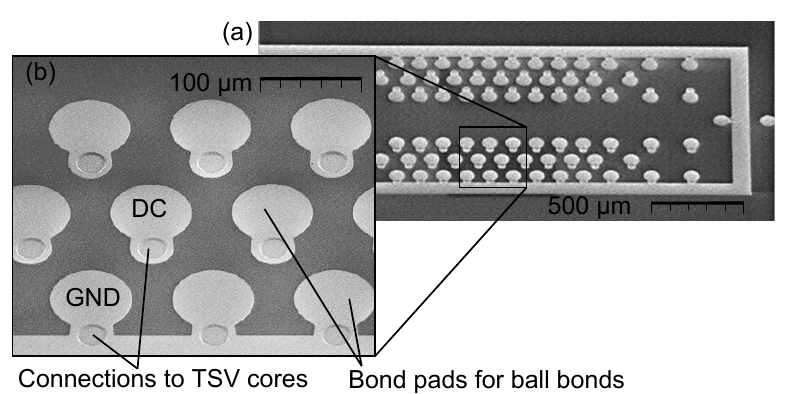}
\caption{\label{fig:BackMetal} (a) Back side of a BGA die; (b) Close-up view.}
\end{figure}

On the back side of the trap die (Fig. \ref{fig:BackMetal}), 48 silicided contacts make electrical connections to the cores of the TSVs and to bond pads; 28 contacts make ground connections to the bulk silicon part of the die. Gold and tin are deposited on the bond pads for gold/tin eutectic solder-bonding to the interposer (Sec. \ref{sec:Assembly}).

\begin{figure}[htbp]
\includegraphics[width=\columnwidth]{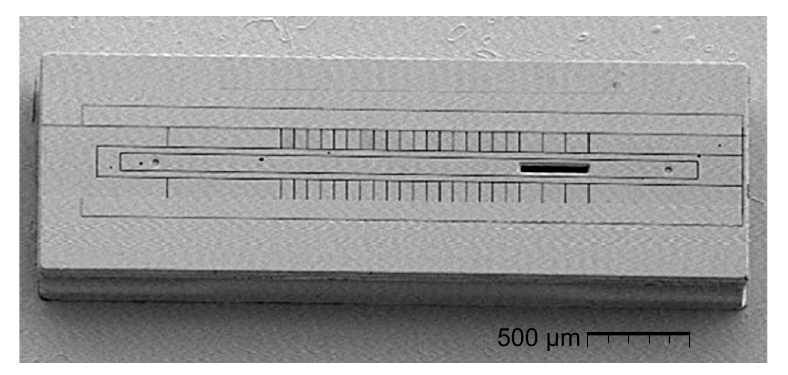}
\caption{\label{fig:TrapDie} Trap die ready to be bonded to an interposer.}
\end{figure}

(v) Making Loading Slots: Slots in the trap die allow passage of a thermal flux of neutral atoms for ion loading (Sec. \ref{sec:TestingOverview}). The loading slot is coated with gold to minimize the potential for charge build-up.  The etch that defines the loading slots also opens up the perimeters of each die so that the die are singulated (separated from one another and from the bulk wafer).  Figure \ref{fig:TrapDie} shows a completed trap die after it has been removed from its carrier wafer.

\subsubsection{\label{sec:Interposer}Interposer Processing}

The interposer is a separate die (1 cm $\times$ 1 cm by 1.3 mm-thick) whose purpose is to elevate the trap die above the rim of the package and to carry electrical signals from the package to the trap die. Signals are routed on two metal layers (Interposer M3 and M2 in Fig. \ref{fig:CrossSection}), each 1.7 $\mu$m Au with Ti and Pt adhesion/barrier layers above and below. Top and bottom ground layers (Interposer M4 and M1), each 1.7 $\mu$m Au, provide electrical shielding. Slots for ion loading are DRIE-etched 300-350 $\mu$m into the front side of the interposer wafer and 1000 $\mu$m in from the back side, meeting to create a continuous opening that aligns with the loading slot in the trap die.  Appendix \ref{app:IntFab} provides details of the interposer fabrication process.

\subsection{\label{sec:Assembly}Assembly}

\begin{figure}[htbp]
\includegraphics[width=\columnwidth]{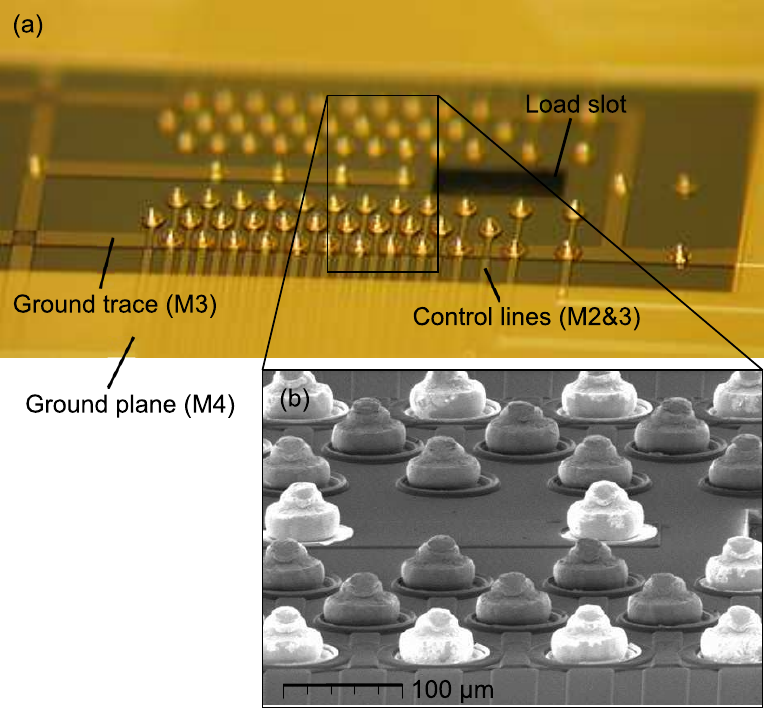}
\caption{\label{fig:AssemblyPhotos} (a) Gold studs attached to the interposer, with interposer metallization layers labeled (see Appendix \ref{app:IntFab}). (b) SEM image; the studs that are attached to ground appear to glow more brighly than those connected to DC electrodes.}
\end{figure}

For final assembly of the ion trap, we bond together the separately fabricated trap die and interposer, then package the combined unit into a CPGA carrier. A conventional gold ball bonder, programmed to leave short bond wires ($\approx 75$ $\mu$m from base to tip), attaches gold studs to the bond pads on top of the interposer (Fig. \ref{fig:AssemblyPhotos}).  With the gold studs as interconnects, the trap die is bonded with a solder-reflow process to the interposer with previously deposited gold and tin layers acting as a gold/tin eutectic solder.

A separate layer of gold studs is placed on the base of the CPGA package, and the interposer (plus trap) is ultrasonically bonded to the CPGA. Wirebonds are then attached from the CPGA to the interposer. One wire bond goes directly from the interposer to the trap die, providing a low-resistance routing for the trap RF signal. The final packaged trap is shown in Figure \ref{fig:Overview}c.

\section{\label{sec:Testing}Testing}

\subsection{\label{sec:TestingOverview}Overview}

\begin{figure}[htbp]
\includegraphics[width=\columnwidth]{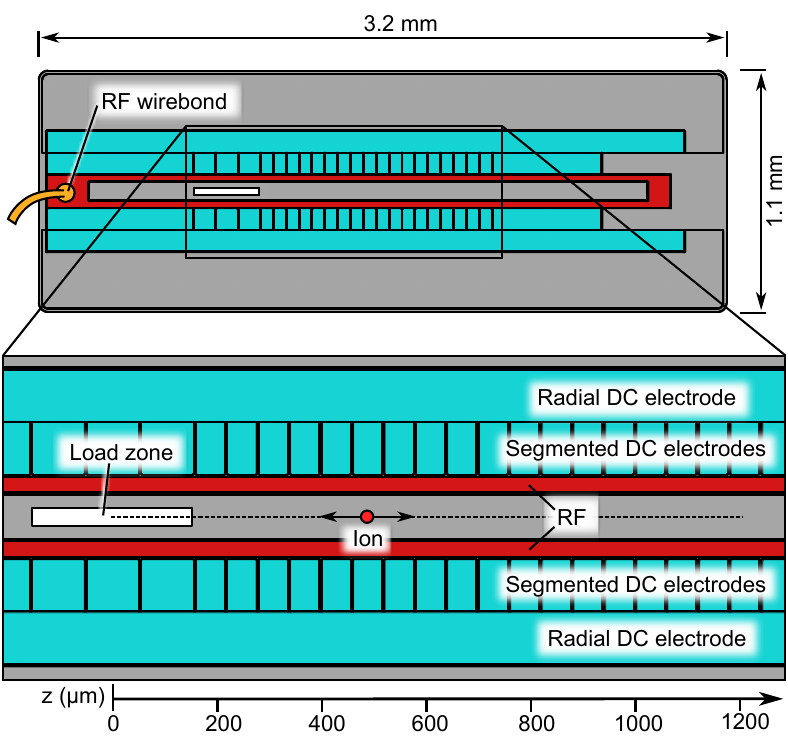}
\caption{\label{fig:BGATrapPlaceholder} Schematic of the BGA trap, showing RF electrodes (red), ion loading slot (white), ground planes (gray), and DC electrodes (teal). Positions are labeled along the trap symmetry axis ($z$).}
\end{figure}

We test the BGA trap by trapping $^{40}$Ca$^+$ ions, using techniques discussed in Refs. \cite{Doret2012,Vittorini2013}.  An oven located below the CPGA carrier supplies neutral Ca atoms which pass through the loading slots in the interposer and trap chip and are photoionized 60 $\mu$m above the trap surface. A potential $V_0 \cos{(\Omega_{RF}t)}$ is applied to the RF electrodes (Fig. \ref{fig:BGATrapPlaceholder}), where $V_0=95$ V and $\Omega_{RF}= 2\pi \times 55.14$ MHz.  This creates a ponderomotive pseudopotential which confines ions radially. Ions are confined axially by applying voltages to the DC electrodes and are transported by varying these voltages to translate the axial potential minimum.

For the following measurements, the axial mode frequency (Sec. \ref{sec:Modes}) is 1.0 MHz and the radial frequencies are 4.2 MHz and 3.7 MHz.  The experiments described below are performed with the ion at $z=488$ $\mu$m (see Fig. \ref{fig:BGATrapPlaceholder}).

\begin{figure}[htbp]
\includegraphics[width=\columnwidth]{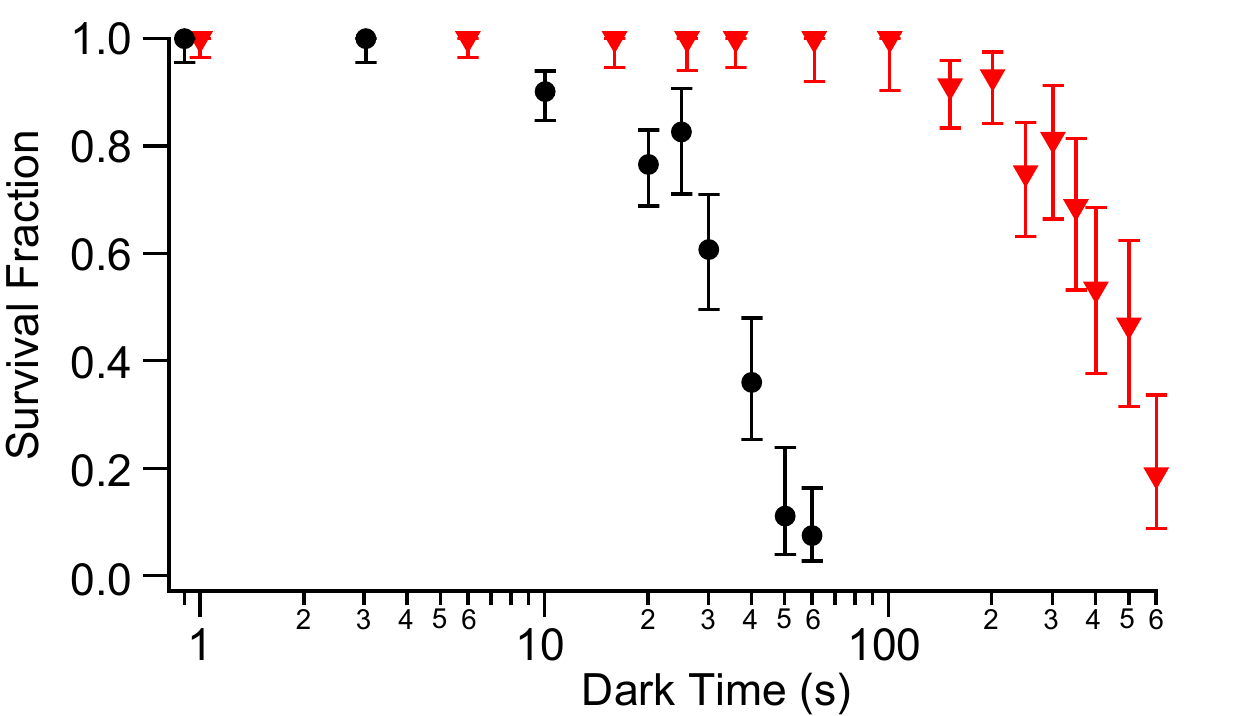}
\caption{\label{fig:BGATwoIonDarkLifetime} Ion lifetime measurement without Doppler cooling. The dark lifetime (50\% survival fraction) is 450 sec for a single ion (red triangles) and 35 sec for two co-trapped ions (black circles).}
\end{figure}

Ions are manipulated with laser light directed parallel to the trap surface.  We use a 397 nm laser tuned near the ${^2S_{1/2}} \rightarrow {^2P_{1/2}}$ transition for Doppler cooling and state detection.  Fluorescence is collected onto a charge-coupled device (CCD) camera and a photomultiplier tube.  An 866 nm laser repumps ions out of the metastable ${^2D_{3/2}}$ level.  With the Doppler cooling beam on, the storage lifetime for a Ca$^+$ ion is several hours.  With the Doppler cooling beam mechanically shuttered, the ion dark lifetime (50\% average survival fraction) is 450 s for a single trapped ion (Fig. \ref{fig:BGATwoIonDarkLifetime}).

\begin{figure}[htbp]
\includegraphics[width=\columnwidth]{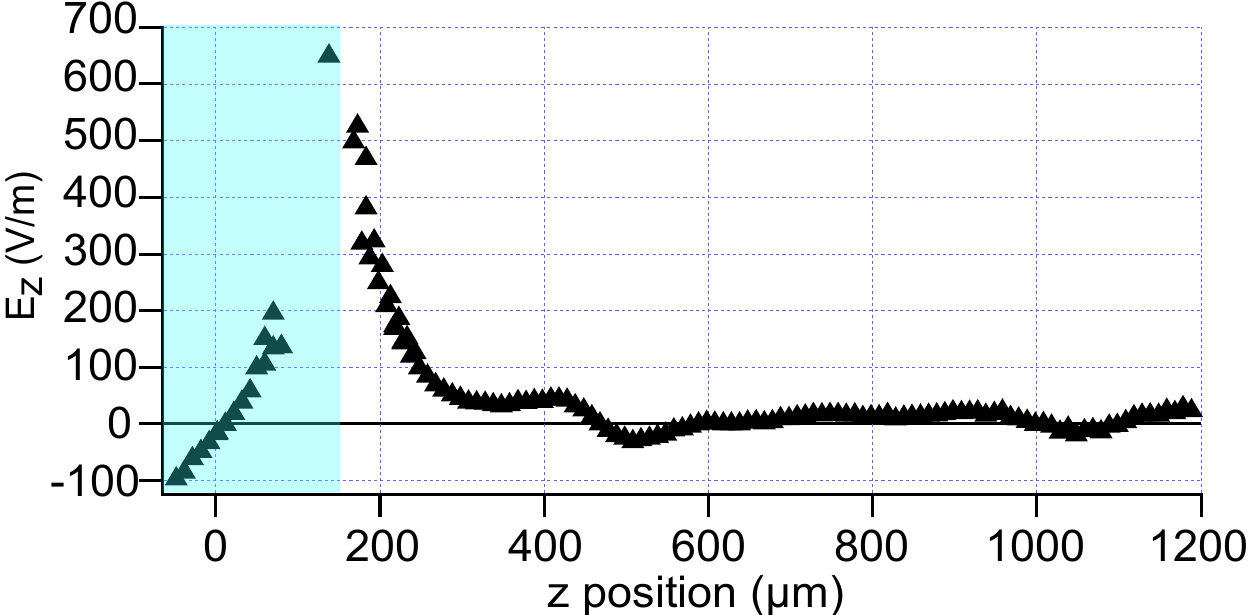}
\caption{\label{fig:FieldMap} Stray axial field $E_z$ measurements over the BGA trap.  Light blue box indicates region of the loading zone.}
\end{figure}

By measuring shifts in the ion equilibrium position as the harmonic trapping potential is varied, we map out stray axial electric field strength $E_z$ over the length of the trap (Fig. \ref{fig:FieldMap}). We are able to trap and transport the ion within a 1.2 mm region.  Stray fields near the load zone edge reach several hundred V/m but drop well below 100 V/m at a distance of 100 $\mu$m from the load slot. These axial stray fields are comparable to those measured in earlier surface-electrode traps \cite{Doret2012,Guise2014,Narayanan2011}.

\subsection{\label{sec:Modes}Motional Modes}

\begin{figure}[htbp]
\includegraphics[width=\columnwidth]{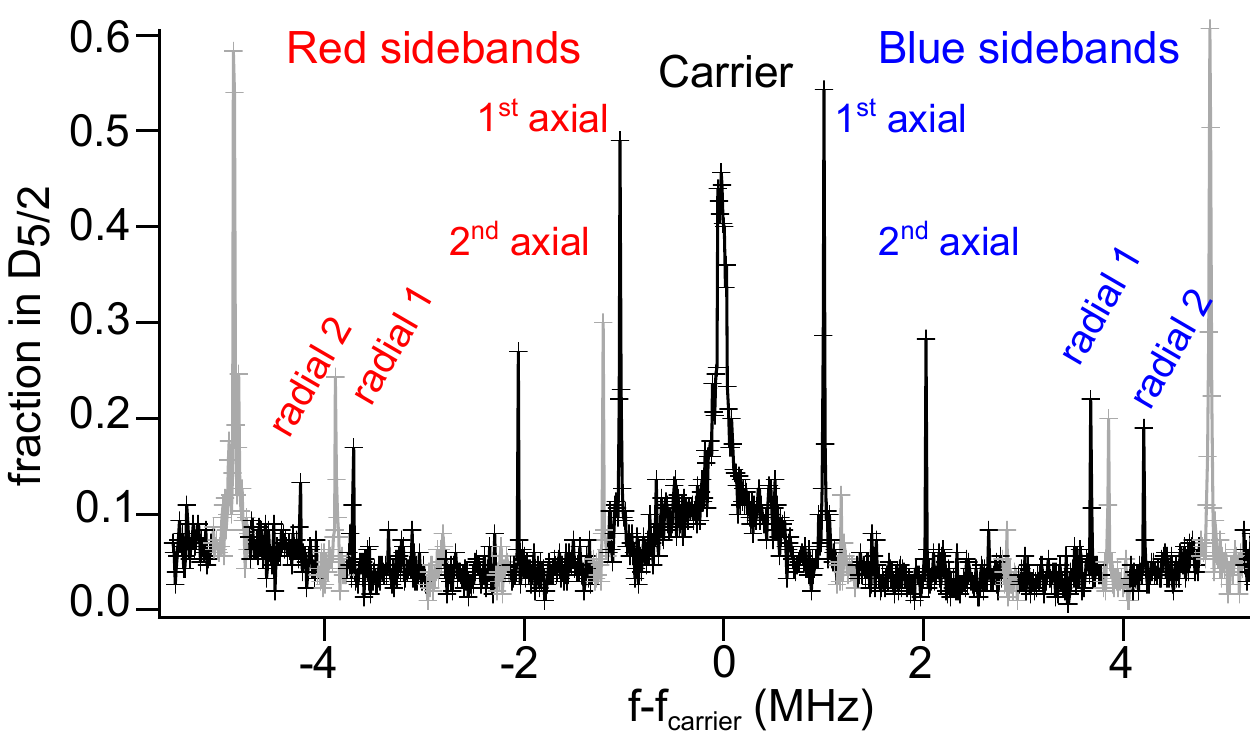}
\caption{\label{fig:ModeScan} Carrier transition and motional sidebands for a trapped Ca$^+$ ion as a function of the 729 nm laser frequency.  Gray peaks are due to adjacent carrier transitions (between different Zeeman sublevels) and their sidebands.}
\end{figure}

To characterize the ion's secular frequencies, we investigate motional sidebands of the trapped ion around a motion-independent carrier frequency.  The chosen carrier is an electric quadrupole transition, ${^2S_{1/2}},m_j=-1/2 \rightarrow {^2D_{5/2}},m_j=-1/2$, with wavelength 729 nm.  We apply a weak magnetic field of $\approx 3$ gauss to split the Zeeman sublevels.  The carrier frequency shifts in direct proportion to magnetic field strength; we use an active compensation coil, driven by feedback from a field probe near the trap, to null out slow B field drift including 60 Hz line noise.  Fig. \ref{fig:ModeScan} shows the spectrum around our carrier transition. Various sidebands appear due to trapped ion motion relative to the driving laser field. Axial sidebands of 1st order ($f_z=1$ MHz) and  2nd order ($f=2f_z$) are clearly resolved along with sidebands at the radial secular frequencies 4.2 MHz and 3.7 MHz.  

\begin{figure}[htbp]
\includegraphics[width=\columnwidth]{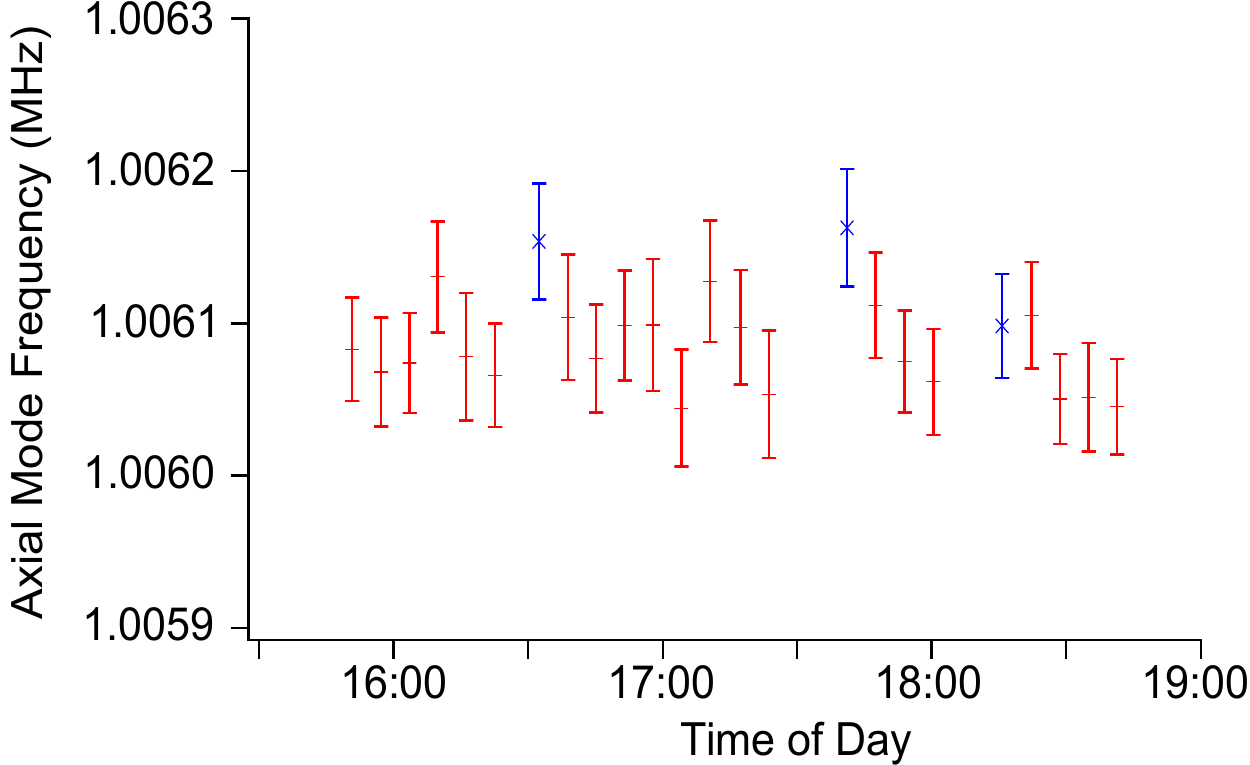}
\caption{\label{fig:ModeStability} Axial mode frequency stability. Blue points indicate measurements following an ion reload event.}
\end{figure}

Axial mode stability is measured by repeatedly scanning over the axial red and blue sideband transitions (labeled ``1st axial'' in Fig. \ref{fig:ModeScan}).  The axial mode frequency remains within a range of 200 Hz (0.02\%) over three hours (Fig. \ref{fig:ModeStability}).  Small shifts of $\approx 100$ Hz (0.01\%) are observed due to ion reload events, which may result from charging of the trap surface by the loading beams. These reload shifts typically decay on a timescale of 5-10 minutes.

\begin{figure}[htbp]
\includegraphics[width=\columnwidth]{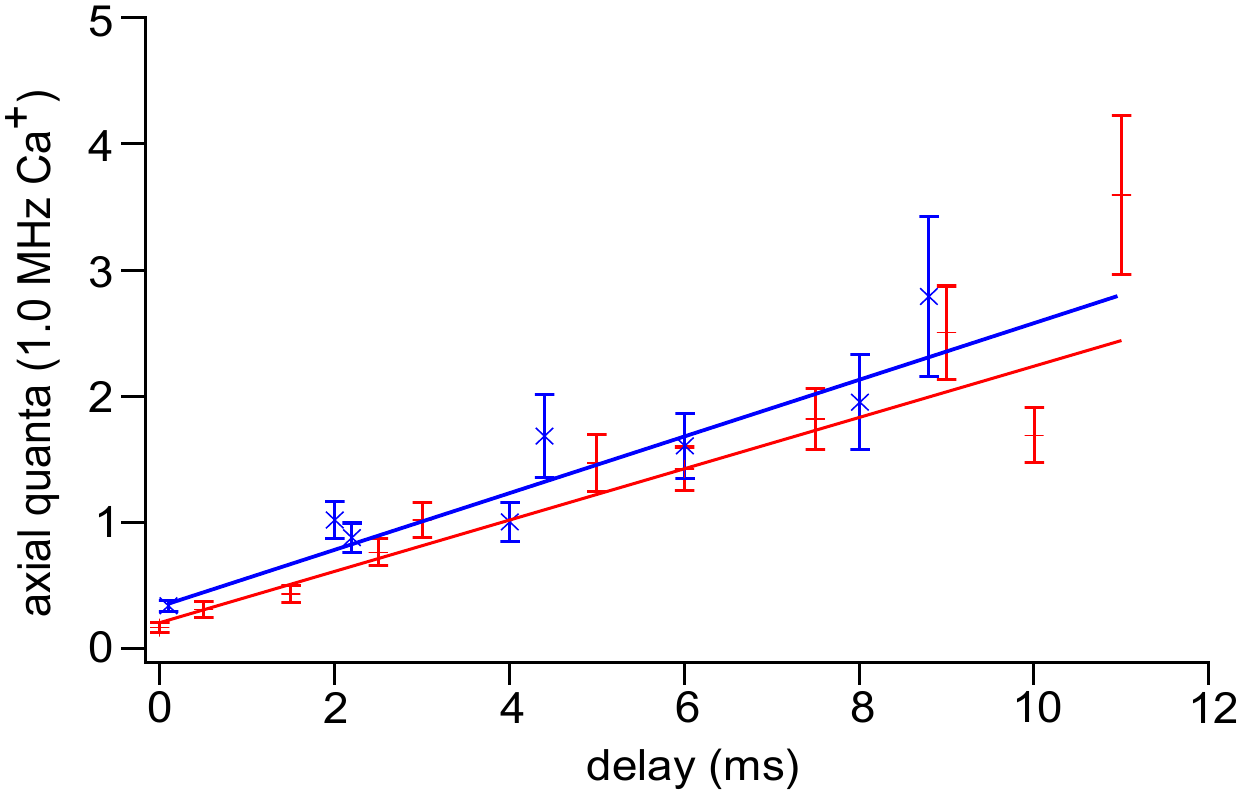}
\caption{\label{fig:HeatingRate} Heating rate of the axial motional mode.  Blue points are with transport at 0.3 m/s during the delay time.  The resulting heating rate is 0.21(1) quanta/ms without transport and 0.22(2) quanta/ms with transport.}
\end{figure}

We use the average phonon occupation number $\bar{n}$ in the axial mode to determine ion heating rates. For low temperatures, $\bar{n}$ may be measured by comparing strengths of the 1st axial red and blue sidebands; i.e. $\bar{n}=x/(1-x)$ where  $x=I_{{red}}/I_{{blue}}$ is the ratio of sideband strengths \cite{Turchette2000}. The ion is sideband cooled to $\bar{n} \approx 0.25$, then allowed to sit without any cooling for a controlled duration.  Following the delay, $\bar{n}$ is remeasured to determine the heating.  Results of this measurement are shown in Fig. \ref{fig:HeatingRate}.  A weighted linear fit to the data gives a heating rate of 0.21(1) quanta/ms.  To check for additional heating due to ion transport, we repeat the experiment with continuous transport of the ion during the delay time.  The transport rate is 0.3 m/s over a 300 $\mu$m region of the trap (round trips between $z=488$ $\mu$m and $z=788$ $\mu$m).  Transport of the ion in this manner does not affect the measured heating rate within our measurement uncertainty (blue points and fit in Fig. \ref{fig:HeatingRate}). The corresponding frequency-scaled electric field noise density, $\omega S_E(\omega) \approx 9 \times 10^{-6}$ V$^2/$m$^2$, is comparable to previous traps with similar ion-electrode separation \cite{Daniilidis2011}.

\subsection{Multiple Ions}

\begin{figure}[htbp]
\includegraphics[width=\columnwidth]{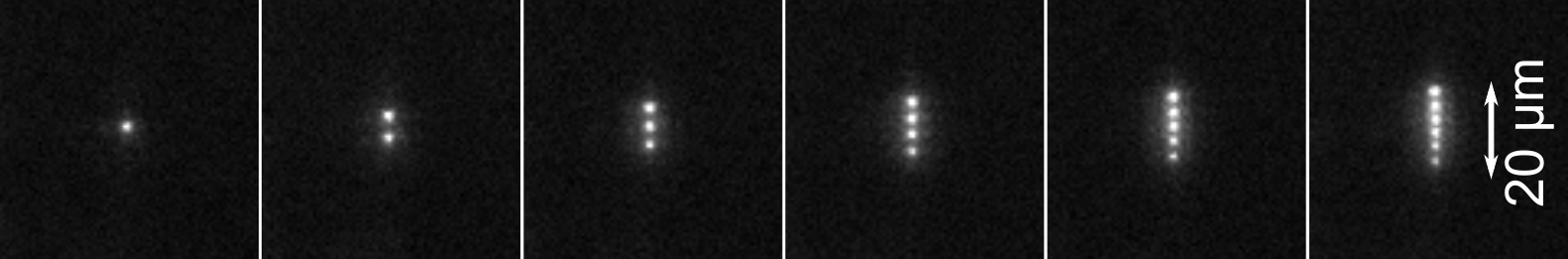}
\caption{\label{fig:MultipleIons} Fluorescence images in the loading zone of one to six trapped ions.}
\end{figure}

We load multiple-ion chains in a single potential well (Fig. \ref{fig:MultipleIons}).  After calibrating transport waveforms to correct for stray electric fields in the $z$ direction (Fig. \ref{fig:FieldMap}), we reliably co-transport multiple ions out of the load slot to our experiment zone at  $z = 488$ $\mu$m.  Transport success rates are near 100\% for co-transport of two or three ions at speeds of 0.33 m/s or slower.
  
\begin{figure}[htbp]
\includegraphics[width=\columnwidth]{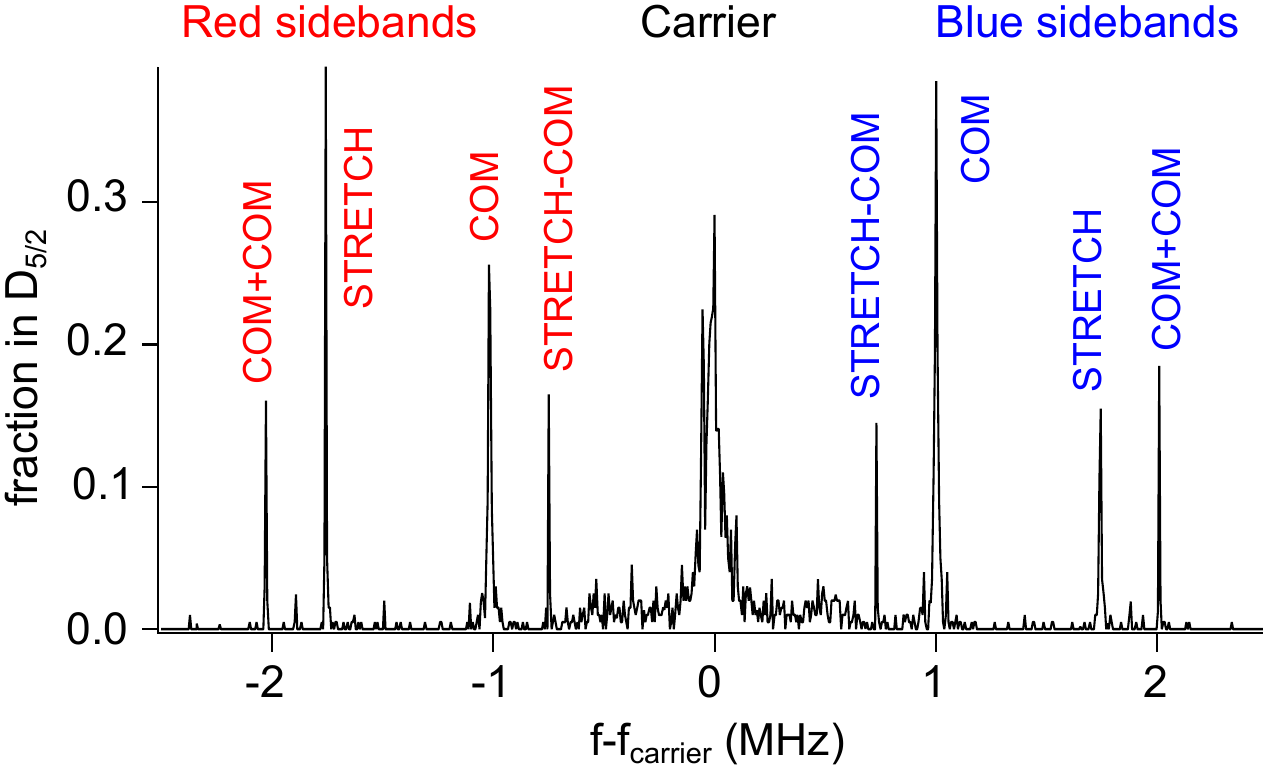}
\caption{\label{fig:TwoIonModes} Carrier transition and motional sidebands for two co-trapped Ca$^+$ ions.}
\end{figure}

In the experiment zone, the two-ion storage lifetime is typically 1-2 hours with Doppler cooling and 35 seconds without (Fig. \ref{fig:BGATwoIonDarkLifetime}).  Note that for a two-ion chain we define lifetime by the loss of either ion.  The two-ion center-of-mass (COM, $f_z=1.0$ MHz) and stretch ($f=1.73$ MHz) axial sidebands are identified in Fig. \ref{fig:TwoIonModes}.  In preparation for two-qubit operations, we sideband cool both axial modes to near the motional ground state, $\bar{n} < 0.5$ quanta in each mode.
 
\subsection{Beam Focusing}

A key advantage of the BGA architecture is the possibility to focus laser beams more tightly on the ion.  Tighter focus of a gate beam can speed up qubit rotation times by increasing intensity at the ion while also reducing crosstalk to neighboring ions.  A focused Gaussian beam will diverge in inverse proportion to the ultimate beam waist; that is, a beam tightly focused above the trap surface will exhibit large angular spread as it enters and exits the trapping region.  The dimensions of the trap chip and carrier thus impose a limit on how tightly we may focus the beam before scattering off some surface enroute.

\begin{figure}[htbp]
\includegraphics[width=\columnwidth]{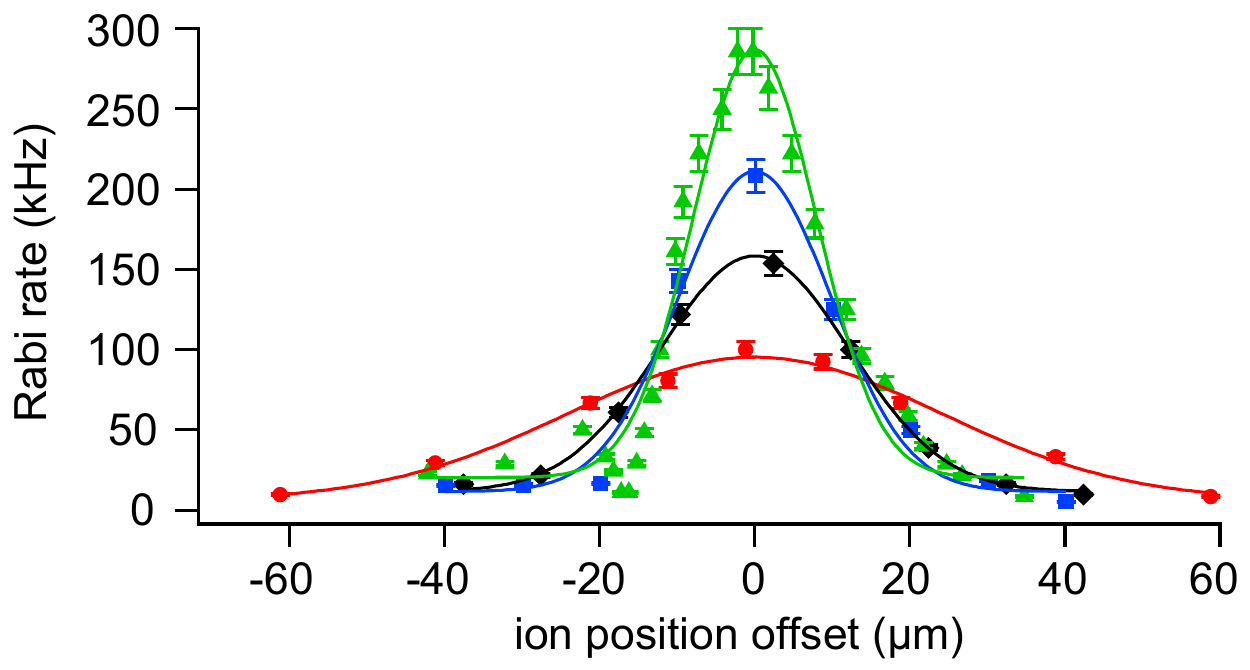}
\caption{\label{fig:BeamWaists} Effect of 729 nm gate beam focus on single qubit transition rate.  Correcting for the 45$^{\circ}$ angle of incidence, the Gaussian fits indicate physical beam waist radii of 24.1 $\mu$m (red), 12.2 $\mu$m (black), 9.9 $\mu$m (blue), and 8.0 $\mu$m (green).}
\end{figure}

We use various lens combinations to focus our 729 nm gate beam onto the trapped ion.  The carrier Rabi rate for single qubit rotations is proportional to the amplitude of the driving electric field (square root of laser intensity) at the ion.  The laser beam profile is measured by translating the ion while holding the 729 nm laser fixed and measuring the associated Rabi frequency; beam waist radius is defined as the $1/e^2$ half-width of intensity profile.   Results for four different beam waists are shown in Fig. \ref{fig:BeamWaists}.  The 729 nm beam is incident at 45$^{\circ}$ to the trap axis, giving rise to a factor of $\sqrt{2}$ between the physical beam waists and the effective beam waists as measured by this technique. 

We model these results by measuring the collimated input beam with a beam profiler, then treating the focusing lenses as ideal optical elements to calculate effective beam waists at the ion. The tightest waist shown in Fig. \ref{fig:BeamWaists} represents the approximate limit of beam focusing over the BGA trap; at 8 $\mu$m waist our model indicates the beam will begin to clip on the CPGA carrier edge, resulting in a loss of a few percent of the beam power (Table \ref{tab:BGAScattering}). Deviations from the Gaussian fit are evident in the wings of the 8 $\mu$m beam (green) in Fig. \ref{fig:BeamWaists}, indicating beam distortion. Beam waists of 10 $\mu$m or larger, meanwhile, are predicted to experience less than 1\% power loss due to clipping. For comparison, the ion separation distance for two co-trapped ions is roughly 6 $\mu$m. As shown in Table \ref{tab:BGAScattering}, such tight beam focusing far exceeds what could be achieved with the same optics in a previous GTRI trap design with larger trap chip area.

\begin{table}
\caption{\label{tab:BGAScattering}Calculated power losses due to clipping of a Gaussian beam off the BGA trap as characterized here and off a GTRI Gen-V trap \cite{Guise2014} if placed at equivalent position in the vacuum chamber.}
\begin{ruledtabular}
\begin{tabular}{lll}
&\multicolumn{2}{c}{Power loss due to beam clipping} \\
$\lambda=729$ nm beam waist&&\\
($\mu$m)&in BGA trap&in Gen-V trap\\
\hline
24.1&$<0.001$\%&6.4\%\\
12.2&0.05\%&21.0\%\\
9.3&0.6\%&26.9\%\\
7.4&3.1\%&32.4\%\\
\end{tabular}
\end{ruledtabular}
\end{table}

\section{\label{sec:MSGate}Entangling gates}

Two qubit entangling gates in Ca$^+$ are highly sensitive to magnetic field noise in the laboratory. Subsequent to the Ca$^+$ testing (Sec. \ref{sec:Testing}), we trapped $^{171}$Yb$^+$ in a second BGA trap (see Refs. \cite{Balzer06_PRA73p041407,Olmschenk07_PRA76p052314} and references therein for details on $^{171}$Yb$^+$).  The $F=0, m_F=0$ to $F=1, m_F=0$ clock transition of $^{171}$Yb$^+$ is magnetic field insensitive at zero field and these states serve as our $\left|0\right>$ and $\left|1\right>$ qubit states, respectively.  We trap pairs of qubits at z = 548 $\mu$m with axial frequency 0.6 MHz and radial frequencies 1.7 MHz and 2.1 MHz. Following the work in Refs. \cite{Hayes10_PRL104p140501,Campbell10_PRL105p090502,Islam14_OptLett39p3238}, we span the hyperfine qubit with a pulsed 355 nm Raman laser. The Raman lasers are counter propagating along the surface of the trap with approximately 10 mW in the first beam at a waist of 15 $\mu$m and 30 mW in the second beam at a waist of 7 $\mu$m. 

\begin{figure}[htbp]
\includegraphics[width=\columnwidth]{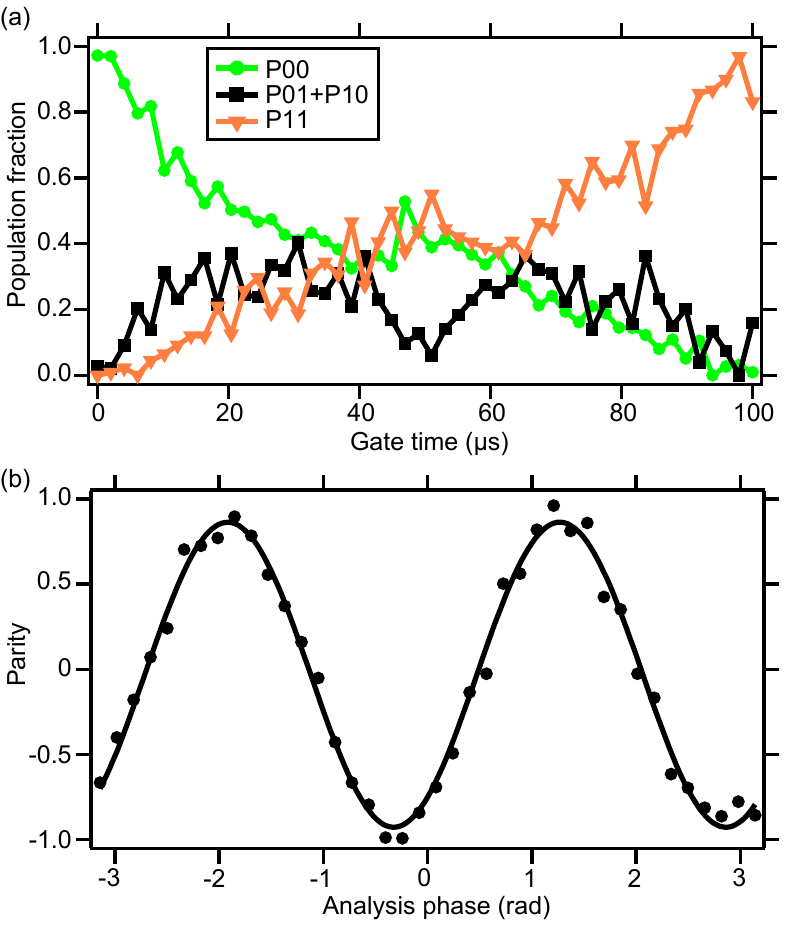}
\caption{\label{fig:MSGate} (a) Population evolution of the $^{171}$Yb$^+$ qubit during the M{\o}lmer-S{\o}rensen gate: P00=both ions in $\left|0\right>$; P01=first ion in $\left|0\right>$, second ion in $\left|1\right>$; P10=first ion in $\left|1\right>$, second ion in $\left|0\right>$; P11=both ions in $\left|1\right>$. (b) Parity measurement (P00+P11-P01-P10) of the entangled ions at gate time of 50 $\mu$s. }
\end{figure}

Starting in the $\left|00\right>$ state, we implement a M{\o}lmer-S{\o}rensen gate \cite{Sorensen00_PRA62p022311} on the 1.7 MHz radial mode to entangle the two ions into the state $(\left|00\right>-i\left|11\right>)/\sqrt{2}$. Figure~\ref{fig:MSGate} shows the population evolution during the entangling gate. It also shows the parity signal at a gate time of 50 $\mu$s as a function of the phase of a subsequent $\pi/2$ pulse on the carrier transition. Following Ref. \cite{Sackett00_Nature404p256}, the resulting two qubit entanglement fidelity is 93(2)\%.

\section{Summary and Conclusions}

A ball-grid array architecture offers significant improvements in size and scalability for microfabricated ion traps. Trench capacitors fabricated into the trap die replace surface filter capacitors, providing a $30\times$ reduction in trap die area over traps with planar capacitors. Through-substrate vias connect the electrodes to pads on the back side of the trap die, eliminating wirebonds from the trap surface. The trap die is bump-bonded to a separate interposer chip for signal routing to a CPGA carrier. Optical access to a trapped ion is improved by the reduced BGA trap chip area and the absence of wirebond obstructions, allowing tighter focusing of laser beams for qubit operations and addressing.

BGA ion trap performance is characterized with $^{40}$Ca$^+$ ions. We measure stray axial electric fields, motional mode stability, heating rate with and without ion transport, and dark lifetimes for one and two ions. By all measures, the BGA trap equals or exceeds results from earlier GTRI surface-electrode traps fabricated with aluminum electrodes and standard wirebonding \cite{Doret2012,Guise2014} and the heating rate is competitive with most microfabricated traps from other groups \cite{Brownnutt2014}. We demonstrate focusing of an addressing 729 nm laser beam to a waist of $\approx 8$ $\mu$m, approaching the separation distance for two co-trapped ions, with minimal power loss due to scattering off trap surfaces. Further improvements in ion addressing could be obtained with cylindrical optics to tighten focus along the trap axis while maintaining beam divergence in the vertical direction relevant to scattering loss. In a second BGA trap, we demonstrate a M{\o}lmer-S{\o}rensen gate with 93\% entanglement fidelity, enabled by tight focusing of 355 nm Raman beams onto a pair of $^{171}$Yb$^+$ ions.

Primary challenges in fabricating the BGA trap involve the trap-to-interposer ball bonding and the TSVs. An ultrasonic process proved unreliable for bonding the trap to the interposer without damaging the die; the solder reflow process described in Sec. {\ref{sec:Assembly} was developed as a robust alternative. The doped silicon TSV process (Sec. {\ref{sec:TrapFab}i) was developed to mitigate risks encountered in fabricating lower-resistance metal vias.  In particular, filling the narrow via holes with plated metal can create voids that trap small amounts of air and degrade the ultra-high vacuum trap environment.

The BGA architecture extends readily to traps of increasing complexity. BGA techniques could simplify fabrication of intricate junction traps, such as the NIST Racetrack \cite{Amini2010} and Sandia MUSIQC Circulator \cite{Monroe2013}, which feature rings of RF rails encircling isolated islands of control electrodes; in such designs it becomes impossible to route leads to all electrodes on a single metal layer. BGA interposer design is flexible and easily modified for signal routing from carriers other than the standard CPGA presented here. The BGA architecture could be integrated with recently demonstrated in-vacuum electronics \cite{Guise2014} to produce ion traps with significantly larger numbers of electrodes at manageable physical size.

\section{ACKNOWLEDGMENTS}
This material is based upon work supported by the Office of the Director of National Intelligence (ODNI), Intelligence Advanced Research Projects Activity (IARPA) under U.S. Army Research Office (ARO) contracts W911NF1210605 and W911NF1010231.  All statements of fact, opinion, or conclusions contained herein are those of the authors and should not be construed as representing the official views or policies of IARPA, the ODNI, or the U.S. Government.

\appendix
\section{\label{app:A}Fabrication Details}

\subsection{\label{app:Metallization}Trap Die Front-side Metallization}

On the trap die, the bottom-most metal layer (M1) is a 1.5 $\mu$m-thick layer of gold surrounded below by a TiW/Pt adhesion/diffusion barrier and above by a Ti adhesion layer. At the locations of TSVs, M1 is patterned into 50 $\mu$m-diameter discs that are separated from the ground plane by 5 $\mu$m gaps.  The discs are connected to the TSVs below through Ohmic contacts and to the above metal layer (M2) by vias. The dielectric between the TSVs and M1 is 1.5 $\mu$m thick; the dielectric between M1 and M2 is 10 $\mu$m thick. Vias in the second dielectric are formed by thermally sloping photoresist and etching the vias in a recipe that transfers the slope into the SiO$_2$.

M2 forms the trap DC and RF electrodes.  M2 is 1.5 $\mu$m of gold with 30 nm of Ti above and below to ensure good adhesion to the SiO$_2$ dielectric.  After M2 has been patterned, CVD SiO$_2$ is deposited over the entire wafer, and a CMP operation is performed to remove most of the topography associated with M2 and the M1-to-M2 vias.  After CMP an additional layer of 1.5 $\mu$m of TEOS SiO$_2$ is deposited.

The last metal layer (M3) is also 1.5 $\mu$m of gold.  A via is etched through the 3.5 $\mu$m SiO$_2$ prior to 3rd metal deposition.  A via plug (also 3.5 $\mu$m thick) is used to fill the via and thereby improve the planarity of the 3rd metal surface. There is minimal print-through into M3 of features other than the via plugs. A temporary layer of Ti and Pt protects M3 during the subsequent oxide etch.
 
\subsection{\label{app:IntFab}Interposer Fabrication}
The process for making the interposer is summarized below, with some detail shown in Figures \ref{fig:CrossSection} and \ref{fig:AssemblyPhotos}.

\begin{enumerate}
\item	Interposer M1 (first metallization layer): M1 forms a uniform ground plane beneath the signal carrying lines excluding the region under the BGA, which is left as bare silicon.  M1 is 1.7 $\mu$m of Au, with Ti and Pt adhesion/barrier layers above and below.  A blanket layer of 4 $\mu$m of SiO$_2$ is deposited above.
\item Interposer M2 (second metallization layer): M2, also 1.7 $\mu$m of Au surrounded by Ti and Pt, carries signals from M3 (see below) to the ball bonds under the BGA. This allows the control signals to pass under ground connections made on M3. A blanket layer of 6 $\mu$m of SiO$_2$ is deposited above.
\item	First via: The first via process etches holes through the SiO$_2$ dielectric down to the metal layer beneath.  In cases where there is a feature on M2 under the via, the etch stops on M2.  Otherwise the etch continues down to M1 (ground).
\item	Interposer M3 (third metallization layer): M3 (1.7 $\mu$m Au) carries signals to M2 from the edges of the interposer and grounds a series of ball bonds on the perimeter of the BGA. The grounded ball bonds provide a low impedance ground path between the BGA and the interposer. A blanket layer of 3 $\mu$m of SiO$_2$ is deposited above.
\item	Second via: The second via process opens up holes though the 3 $\mu$m of SiO$_2$ down to M3.  Holes near the perimeter of the interposer expose pads on M3 for wire-bonding to the package.  Holes near the center of the interposer expose pads on M3 for solder-bonding to the trap die.  
\item	Interposer M4 (fourth metallization layer): M4 (1.7 $\mu$m Au) is a ground plane covering the bulk of the die except for the region under the BGA.
\item	Bond pads: An additional layer of metal is deposited, with thickness and properties optimized for bonding.
\item	Front-side loading slots: Holes for the ion loading slots are DRIE-etched 300-350 $\mu$m into the front side of the wafer.
\item	Back-side loading slots:  Holes for the ion loading slots are DRIE-etched 1000 $\mu$m in from the back side of the wafer, meeting the front-side loading slot holes 300 $\mu$m from the front surface of the interposer to create a continuous opening.
\end{enumerate}

\end{document}